\documentclass{emulateapj}

\shorttitle{Stratification of sunspot umbral dots}
\shortauthors{Riethm\"uller et al.}

\begin{document}

\title{Stratification of sunspot umbral dots from inversion of Stokes profiles recorded by $Hinode$}

\author{T. L. Riethm\"uller, S. K. Solanki, and A. Lagg}
\affil{Max-Planck-Institut f\"ur Sonnensystemforschung, Max-Planck-Str. 2, 37191 Katlenburg-Lindau, Germany;}
\email{[riethmueller;solanki;lagg]@mps.mpg.de}

\begin{abstract}
   This work aims to constrain the physical nature of umbral dots (UDs) using high-resolution
   spectropolarimetry. Full Stokes spectra recorded by the spectropolarimeter on $Hinode$ of 51 UDs
   in a sunspot close to the disk center are analyzed. The height dependence of the temperature,
   magnetic field vector, and line-of-sight velocity across each UD is obtained from an inversion
   of the Stokes vectors of the two Fe~I lines at 630~nm. No difference is found at higher altitudes
   ($-3 \le \log(\tau_{500}) \le -2$) between the UDs and the diffuse umbral background. Below that level
   the difference rapidly increases, so that at the continuum formation level ($\log(\tau_{500}) = 0$)
   we find on average a temperature enhancement of 570~K, a magnetic field weakening of 510~G, and
   upflows of 800~m~s$^{-1}$ for peripheral UDs, whereas central UDs display an excess temperature of on
   average 550~K, a field weakening of 480~G, and no significant upflows. The results for, in particular,
   the peripheral UDs, including cuts of magnetic vector and velocity through them, look remarkably similar
   to the output of recent radiation MHD simulations. They strongly suggest that UDs are produced by
   convective upwellings.
\end{abstract}


%__________________________________________________________________
\section{Introduction}

   The energy transport immediately below the solar surface is mainly determined by convective
   processes that are visible as granulation patterns in white-light images of the quiet photosphere.
   This convection is suppressed inside sunspot umbrae due to the strong vertical magnetic field, but some
   form of magnetoconvection \citep{Weiss2002} is needed to explain the observed umbral brightnesses.
   Umbral fine structure such as light bridges or umbral dots, dotlike bright features inside umbrae,
   may well be manifestations of magnetoconvection. Different models have been proposed to explain UDs,
   e.g., columns of field-free hot gas in between a bundle of thin magnetic flux ropes
   \citep{Parker1979,Choudhury1986}, or spatially modulated oscillations in a strong magnetic field
   \citep{Weiss1990}. Recent numerical simulations of three-dimensional radiative magnetoconvection
   \citep{Schuessler2006} reveal convective plumes that penetrate through the solar surface and look very
   much like UDs. Although recent broadband images may have spatially resolved UDs
   \citep[][in preparation]{Sobotka2005,Riethmueller2008}, spectropolarimetry is needed to learn more about their physical nature.
   Previous spectroscopic observations led to heterogeneous results. \citet{Kneer1973} found that
   UDs exhibit upflows of 3~km~s$^{-1}$ and a 50\% weaker magnetic field compared to the nearby umbra,
   whereas \citet{Lites1991} and \citet{Tritschler1997} reported
   little field weakening. Finally, \citet{SocasNavarro2004} observed a weakening
   of 500~G and upflows of a few 100~m~s$^{-1}$. More details can be found in the reviews of umbral fine
   structure by \citet{Solanki2003} and \citet{Sobotka2006}. One reason for the
   difference in results has been the influence of scattered light and variable seeing, which affect
   the different analyzed data sets to varying degrees. It therefore seems worthwhile to invert
   Stokes profiles obtained by the spectropolarimeter (SP) on the $Hinode$ spacecraft. The usefulness
   of $Hinode$ data for the study of UDs was demonstrated by \citet{Bharti2007}, who found
   that large UDs show dark lanes whose existence had been predicted by \citet{Schuessler2006}.

%__________________________________________________________________
\section{Observations and data reduction}

   The data employed here were acquired by the spectropolarimeter \citep{Lites2001}
   of the Solar Optical Telescope \citep[SOT,][]{Suematsu2008} onboard $Hinode$. They are
   composed of full Stokes spectra in the Fe~I line pair around 6302~{\AA} and the nearby continuum
   of a sunspot of NOAA AR~10933 recorded from 12:43 to 12:59~UT on 2007 January 5
   using the 0.16$^{\prime\prime}$x164$^{\prime\prime}$ slit. At this time the sunspot was located at
   a heliocentric angle of 4$^\circ$, i.e. very close to disk center. The observations covered the
   spectral range from 6300.89 to 6303.26~{\AA}, with a sampling of 21~{m\AA}~pixel$^{-1}$. The SP
   was operated in its normal map mode, i.e. both the sampling along the slit and the slit-scan
   sampling were 0.16$^{\prime\prime}$, so that the spatial resolution should be close to the
   diffraction limit of $1.22~\lambda/D = 0.32^{\prime\prime}$. The integration time per slit position
   was 4.8~s which reduced the noise level to $10^{-3}~I_c$.

\begin{figure}
   \centering
   \includegraphics[width=\linewidth]{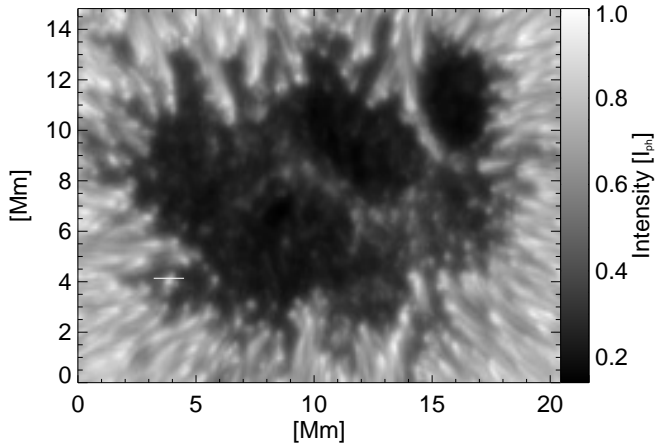}
   \caption{Continuum intensity map of the sunspot NOAA~10933 as observed by the $Hinode$ SOT/SP on 2007
   January 5. Heliocentric angle is $\theta$~=~4$^\circ$. Intensities are normalized to the intensity
   level of the quiet photosphere $I_{ph}$. The white line at (4,4)~Mm marks the cut through an umbral
   dot (UD) that is discussed in greater detail.}
   \label{FigUmbra}
   \end{figure}

   The data were corrected for dark current, flat field, and instrumental polarization with the help of the
   SolarSoft package.\footnote{See \url{http://www.lmsal.com/solarsoft}.} A continuum intensity image (put together from the
   slit scan) of the chosen umbra is shown in Figure~\ref{FigUmbra}. Due to the large slit length
   we are always able to find a sufficiently extensive region of quiet Sun that is used to normalize
   intensities.

%________________________________________________________________
\section{Data analysis}

   To obtain atmospheric stratifications of temperature ($T$), magnetic field strength ($B$), and line-of-sight
   velocity ($v_{LOS}$) we use the inversion code SPINOR described by \citet{Frutiger2000b}.
   This code incorporates the STOPRO routines \citep{Solanki1987}, which compute synthetic Stokes
   profiles of one or more lines upon input of their atomic data and one or more model atmospheres.
   Local thermodynamic equilibrium conditions are assumed and the Unno-Rachkovsky radiative transfer
   equations are solved. The inversions use an optical depth scale as the appropriate coordinate
   for radiative transfer problems. For reasons of comparability we use the optical depth at 500~nm
   ($\tau_{500}$). Starting with an initial guess model, the synthetic profiles were iteratively
   fitted to observed data using response functions (RFs) and the merit function $\chi^2$
   \citep{RuizCobo1992,Frutiger2000a} is minimized. With the help of the  RFs we find that the
   Fe~I line pair at 6302~{\AA} is mainly formed within the $\log(\tau_{500})$ interval [$-3,0$], which
   corresponds to a height range of about 400~km under hydrostatic equilibrium conditions in the umbra.
   The free parameters are defined at the four nodes $-3$, $-2$, $-1$, and $0$ of the $\log(\tau_{500})$ grid.
   The atmospheric stratification is then interpolated using splines onto a 10 times finer $\log(\tau_{500})$
   grid.

   The first step of our analysis is the wavelength calibration required to determine line-of-sight
   (LOS) velocities. For every slit position we average the Stokes $I$ profiles of all locations
   along the slit whose total polarization $P = \int(Q^2 + U^2 + V^2)^{1/2}d\lambda$ is negligible, since
   those locations are assumed to represent the quiet Sun. This mean $I$ profile is used to fit Voigt profiles
   to the two Fe~I lines from which the line center wavelengths are determined. The convective blueshift
   of 140~m~s$^{-1}$ \citep[see][]{MartinezPillet1997,Dravins1981} is then removed.

   The next step is to find an appropriate model atmosphere. Since we are interested in the atmospheric
   stratification of temperature, magnetic field strength, and LOS velocity within a UD, these
   three atmospheric parameters are assumed to be height dependent, whereas field inclination and azimuth angle,
   micro-turbulence, and macro-turbulence are assumed to be height independent. We experimented intensively
   with adding a second model component to represent the stray light, but the inversion results did not
   improve significantly, confirming the almost negligible stray light in the SP. Therefore, in the
   interests of a robust inversion, we forbore from adding a stray light component, thus reducing the number
   of free parameters.

   Lastly, we have to find initial guesses for all free parameters. We use an initial temperature stratification
   according to the umbral core model L of \citet{Maltby1986} and assume a vertical magnetic
   field of 2000~G and zero LOS velocity at all heights. Initial guesses for micro-turbulence and macro-turbulence
   are 0.1 and 2~km~s$^{-1}$, respectively. Other initial guesses gave very similar results, except for
   a limited number of outliers. For these, repeating the inversion with an initial guess close to the final
   result of one of the neighboring pixels returned values consistent with those obtained for the other pixels.

%________________________________________________________________
\section{Inversion results}

\begin{figure}
   \centering
   \includegraphics[width=\linewidth]{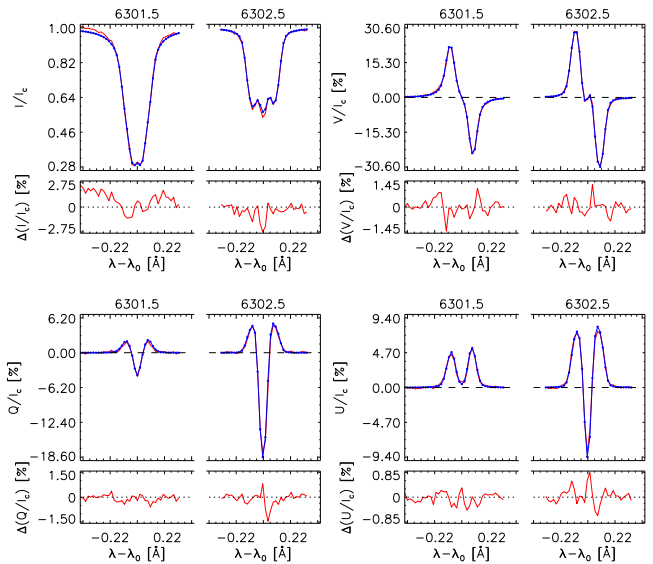}
   \caption{Stokes $I$, $V$, $Q$ and $U$ profiles from the center of the UD marked in Figure~\ref{FigUmbra}.
   Red lines are the measured, blue lines the best-fit profiles, i.e. the inversion result. The bottom parts
   of each panel show the difference between the two on an expanded scale.}
   \label{FigProfileUdBright}
   \end{figure}

\begin{figure}
   \centering
   \includegraphics[width=\linewidth]{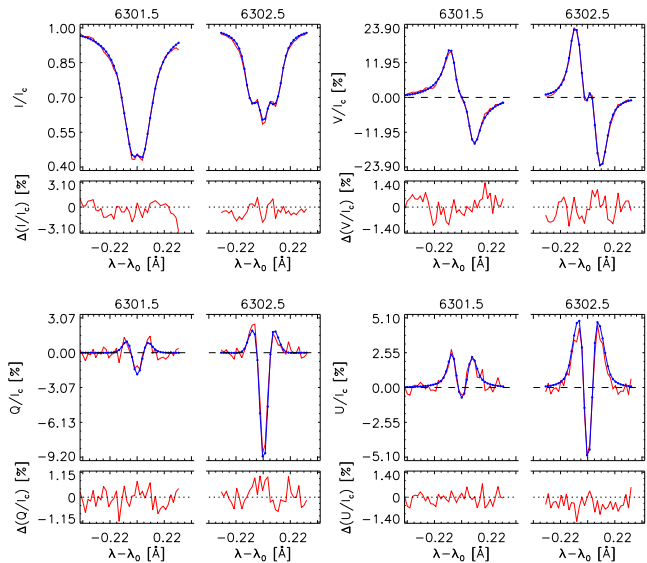}
   \caption{The same as Figure~\ref{FigProfileUdBright}, but for Stokes $I$, $V$, $Q$ and $U$ profiles of
   the diffuse background near the UD.}
   \label{FigProfileUdDark}
   \end{figure}

   We analyzed a total of 51 UDs, which were identified by applying the multilevel tracking (MLT) algorithm
   \citep[][in preparation]{Bovelet2001,Riethmueller2008}. For each UD
   the location of its core was identified, a cut was made through it, reaching to the neighboring diffuse
   background (DB), and the profiles from all the pixels along this cut were inverted. We first discuss the
   results for the UD marked in Figure~\ref{FigUmbra}, chosen because of its brightness, which leads
   to particularly small error bars. A comparison of the measured profiles with the best-fit profiles resulted
   from the inversion can be seen in Figure~\ref{FigProfileUdBright} for the UD and in Figure~\ref{FigProfileUdDark}
   for the DB selected as the location of lowest continuum intensity in a 1.4~$\times$~1.4~Mm$^2$ environment
   of the UD center. Due to the low signal in the dark background the measured DB profiles are much noisier
   than the UD center's profiles, but in general, the Stokes spectra can be fitted remarkably well.

\begin{figure}
   \centering
   \includegraphics[width=\linewidth]{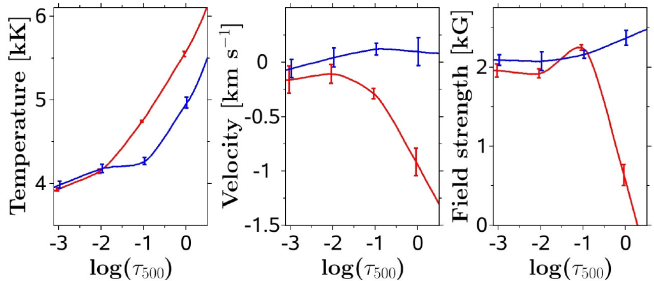}
   \caption{Atmospheric stratification obtained from the Stokes profiles at the location of the UD's center
   (red lines) and from the Stokes profiles of the diffuse background near the UD (blue lines).
   The formal errors of the inversion at the used optical depth nodes are indicated by bars. Negative
   LOS velocity values indicate upflows.}
   \label{FigSingleAtm}
   \end{figure}

   The stratification of the retrieved atmospheric parameters $T$, $v_{LOS}$, and $B$ in the center of the UD
   and in the DB are plotted in Figure~\ref{FigSingleAtm}. In the upper photosphere ($-3 \le \log(\tau_{500})
   \le -2$) the error bars overlap; i.e. we find little significant difference between UD and DB. In the
   deeper photosphere, however, the inversions return strongly different stratifications. Thus, the UD
   temperature is higher than the DB temperature, consistent with the intensity enhancement of the UD in
   the continuum map. The LOS velocity (which is identical to the vertical velocity due to the small
   heliocentric angle) exhibits strong upflows in the UD center, whereas the DB is nearly at rest.
   The magnetic field strength is roughly 2~kG for the heights $-3 \le \log(\tau_{500}) \le -1$. Below
   $\log(\tau_{500}) = -1$ the field strength of the UD decreases strongly with depth, whereas the field
   strength of the DB increases moderately.

\begin{figure}
   \centering
   \includegraphics[width=\linewidth]{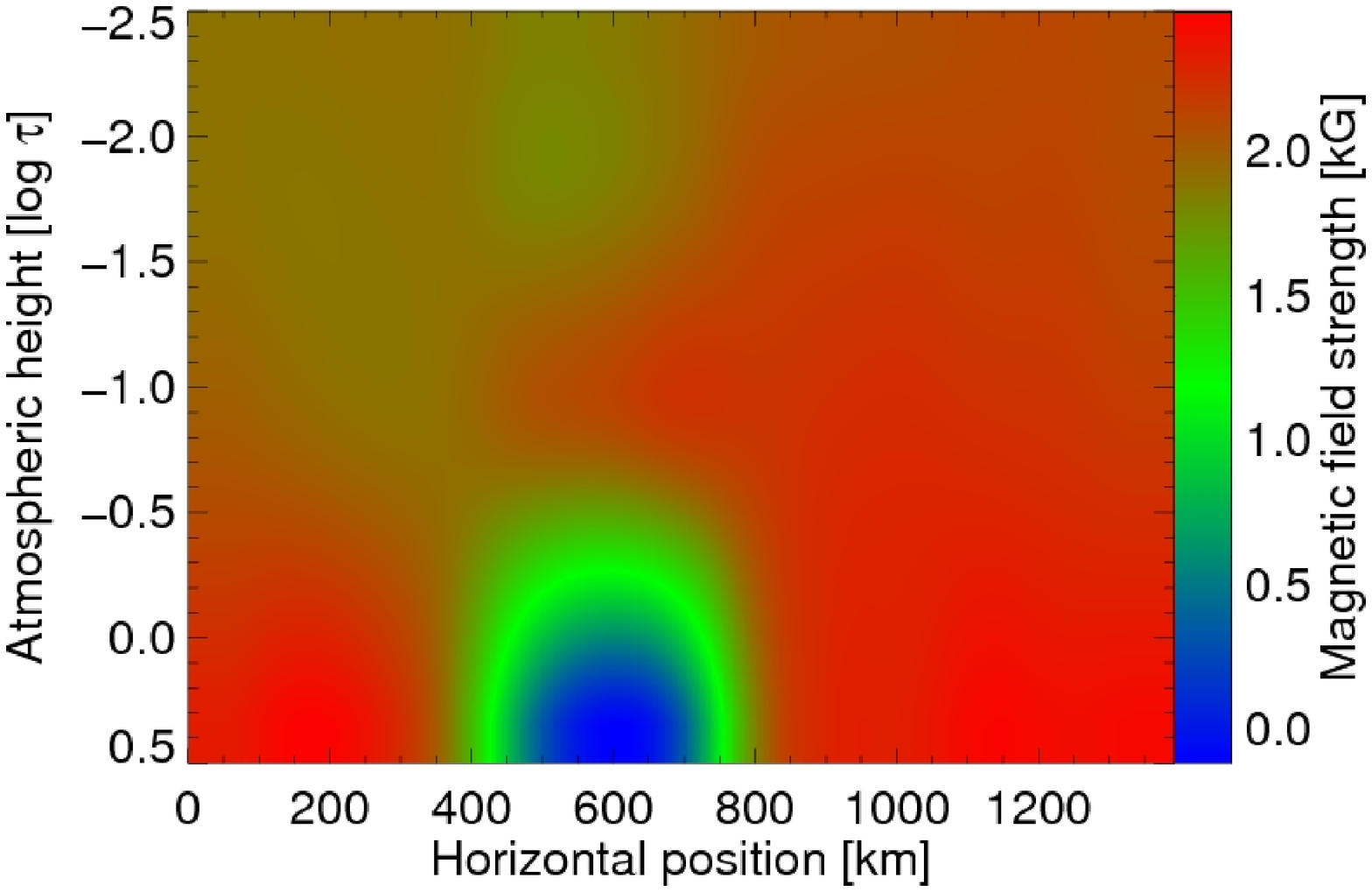}\\
   \includegraphics[width=\linewidth]{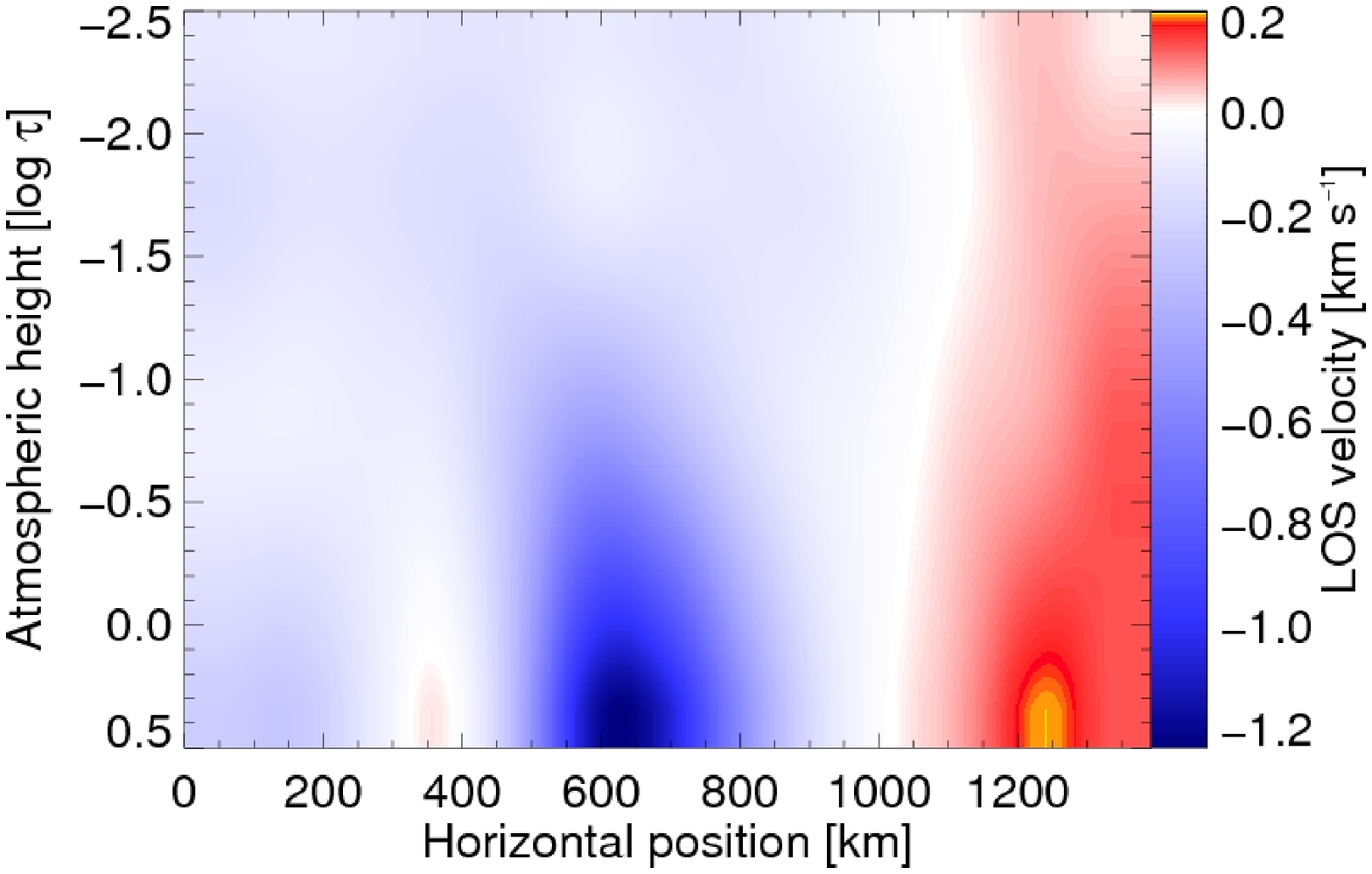}
   \caption{Vertical cut through the UD marked in Figure~\ref{FigUmbra} in the direction indicated by the
   white line. Colors of the top panel indicate magnetic field strength. The bottom panel shows LOS velocity.
   Negative velocities are upflows.}
   \label{FigVerticalCut}
   \end{figure}

   The vertical cuts of magnetic field strength and LOS velocity through 13 pixels lying along the white
   line in Figure~\ref{FigUmbra} are shown in Figure~\ref{FigVerticalCut}. Jumps from one pixel to the next
   were smoothed through interpolation. There is clear evidence for a localized decrease in UD field
   strength in the low photosphere, co-located with an upflow that extends higher into the atmosphere and
   a weak downflow on at least one side. The magnetic fields are 4$^\circ$ more inclined in the UD than
   they are in the DB around the UD. Figure~\ref{FigVerticalCut} looks remarkably like Fig.~2 of
   \citet{Schuessler2006}, in spite of the fact that Figure~\ref{FigVerticalCut}
   is plotted on an optical depth scale in the vertical direction and is thus distorted by an unknown
   amount relative to a corresponding figure on a geometrical scale.

\begin{figure}
   \centering
   \includegraphics[width=\linewidth]{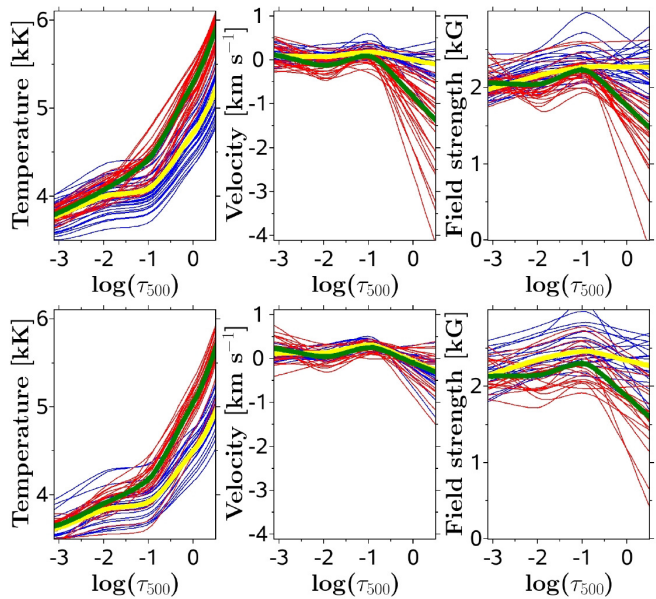}
   \caption{Atmospheric stratifications of peripheral umbral dots (top 3 panels) and central umbral dots
   (bottom 3 panels). The red lines show the stratification at the location of the UD's center and the
   blue lines correspond to the nearby diffuse background. The green line is the weighted average of all
   red lines and the yellow line is the weighted average of all blue lines, where we used the reciprocal
   error bars as weighting factors.}
   \label{FigMultipleAtms}
   \end{figure}

   Next we discuss all 51 analyzed UDs. In the literature we often find a separation into two UD regimes.
   For example, \citet{Grossmann1986} differentiate between peripheral
   UDs (PUDs) and central UDs (CUDs), i.e. between UDs that are born close to the umbra-penumbra boundary
   and UDs that are born deep in the umbra. We follow this distinction and plot the obtained stratifications
   of the 30 PUDs (distance to umbra-penumbra boundary less than 2000~km) in the top panels of
   Figure~\ref{FigMultipleAtms}, while the remaining 21 CUDs are represented in the bottom panels of
   Figure~\ref{FigMultipleAtms}. The results largely mirror those obtained for the UD discussed above. In the
   upper atmosphere UDs center and DB do not differ in their mean values of $T$, $v_{LOS}$, and $B$.
   On average, the CUDs are about 150~K cooler than the PUDs in the upper atmosphere, just as the DB
   around the CUDs is cooler than the DB around the PUDs. At $\log(\tau_{500}) = 0$ we find that PUDs are
   570~K hotter than the local DB and CUDs are 550~K hotter than the DB in their vicinity. The magnetic
   field strength at $\log(\tau_{500}) = 0$ is weakened by about 510~G for PUDs and 480~G for CUDs, whereas
   only PUDs exhibit significant upflows of about 800 m~s$^{-1}$. The mean LOS velocity shows no difference between
   CUD centers and DB. In order to make sure that an upflow is not being missed due to the lower S/N ratio
   of the CUD Stokes profiles, we have also averaged the Stokes profiles of all the CUDs. An inversion
   of there averaged Stokes profiles gave a result that agrees with the averaged stratifications ($green~line$)
   in the bottom panels of Figure~\ref{FigMultipleAtms} within the error bars. This suggests that any upflow
   velocity in CUDs is mostly restricted to layers below the surface or is too concentrated or too weak to be
   detected by the inversions. Finally, we find that the magnetic field of the PUDs is on average
   4$^\circ$ more horizontal than for their DB. We see no inclination difference for CUDs.

%______________________________________________________________

\section{Discussion}

   We identified 30 peripheral and 21 central umbral dots in $Hinode$ spectropolarimetric data of a sunspot
   within 4$^\circ$ of disk center. With the help of Stokes profile inversions of the Fe~I lines at
   630~nm we determined the stratifications of temperature, magnetic field strength, and LOS velocity.
   The present work differs from that of \citet{SocasNavarro2004} in the superior quality
   of the employed data with twice the spatial resolution and practically no scattered light. This allows
   a detailed determination of the atmospheric stratification. The higher spatial resolution of the
   $Hinode$ SP data also allows us to, for the first time, reconstruct both the horizontal and the vertical
   structure of UDs. We also extended the analysis to a more numerous statistical ensemble of 51 UDs.

   Vertical cuts through UDs provide a remarkable confirmation of the results of MHD simulations of
   \citet{Schuessler2006}: both show that UDs differ from their surroundings
   mainly in the lowest visible layers, where the temperature is enhanced and the magnetic field is
   weakened. We found a temperature enhancement of 550~K and a magnetic field reduction of about 500~G
   (at optical depth unity). In addition, PUDs display upflow velocities of 800~m~s$^{-1}$ on average,
   again in good agreement with the simulations. There are also some differences between our results
   and those of \citet{Schuessler2006}. Thus, according to our inversions the
   magnetic field strength of the DB is somewhat depth dependent. This was not the case for the MHD
   simulations due to the used periodic boundary conditions. Furthermore, although some of the UDs
   display a weak downflow bounding the strong central upflow (see Figure~\ref{FigVerticalCut}),
   these are neither as narrow nor as strong as the downflows at the ends of dark lanes as reported
   by \citet{Schuessler2006}, probably due to the limited spatial resolution of our data. We may also
   be missing some of the narrow downflows by considering only single cuts across individual UDs.

   \citet{SocasNavarro2004} reported 10$^\circ$ more inclined magnetic fields in PUDs.
   This result is qualitatively confirmed by our work; we find an inclination increase of 4$^\circ$ for PUDs
   but no increase for CUDs, which can be assumed as a further hint that the main part of the CUD structure
   is below the surface. These results can be interpreted in terms of the strong DB fields expanding with
   height and closing over the UD, as proposed by \citet{SocasNavarro2004}.

%______________________________________________________________

\end{document}